\documentclass[pra,aps,superscriptaddress,amsmath,amssymb,twocolumn,reprint]{revtex4-2}
\usepackage{CJKutf8}
\usepackage{graphicx}
\usepackage{subfigure}
\usepackage{dcolumn}
\usepackage{bm,braket}
\usepackage{mathptmx}
\usepackage{amsmath}
\usepackage{amsfonts}
\usepackage{amssymb}
\usepackage{mathrsfs}
\usepackage{float}
\usepackage{verbatim}
\usepackage{xcolor}
\usepackage[colorlinks=true,linkcolor=blue,anchorcolor=red,citecolor=blue, urlcolor=blue]{hyperref}
\usepackage{ulem}
\usepackage{pifont}

\makeatletter
\renewcommand*\env@matrix[1][\arraystretch]{%
	\edef\arraystretch{#1}%
	\hskip -\arraycolsep
	\let\@ifnextchar\new@ifnextchar
	\array{*\c@MaxMatrixCols c}}
\makeatother

\def\be{\begin{equation}}
	\def\ee{\end{equation}}

\newlength{\seplinewidth}
\newlength{\seplinesep}
\colorlet{sepline}{orange}

\begin{document}
	\begin{CJK*}{UTF8}{gbsn}
		\title{Quantum Dynamical Tunneling Breaks Classical Conserved Quantities}
		
		\author{Lingchii Kong(孔令琦)}
		\affiliation{International Center for Quantum Materials, School of Physics, Peking University, Beijing 100871, China}
		\author{Zongping Gong(龚宗平)}
		\affiliation{Department of Applied Physics, University of Tokyo, 7-3-1 Hongo, Bunkyo-ku, Tokyo 113-8656, Japan}
		\author{Biao Wu(吴飙)}%
		\email{wubiao@pku.edu.cn}
		\affiliation{International Center for Quantum Materials, School of Physics, Peking University, Beijing 100871, China}
		\affiliation{
			Wilczek Quantum Center, School of Physics and Astronomy, Shanghai Jiao Tong University, Shanghai 200240, China}
		
		\date{\today}		
		
		\begin{abstract}
	We discover that  quantum dynamical tunneling, 
	occurring between phase space regions 
	in a classically forbidden way, can break conserved quantities in pseudointegrable systems. 
	We rigorously prove that a conserved quantity in a class of 
	typical pseudointegrable systems 	can be broken quantum mechanically. 
	We then numerically compute the uncertainties of this broken conserved quantity, which remain non-zero for up to $10^5$ eigenstates and exhibit universal distributions similar to energy level statistics. Furthermore, all the eigenstates with large uncertainties show the superpositions of regular orbits with different values of the conserved quantity, showing definitive manifestation of dynamical tunneling. A random matrix model is constructed to successfully reproduce the level statistics in pseudointegrable systems.
			
		\end{abstract}
		
		\maketitle
		
		\section{Introduction} 
		Dynamical tunneling is a fundamental quantum phenomenon which refers to classically forbidden transitions between phase-space regions, even without an energy barrier in the middle \cite{Davis1981}. It has been well stuided in both classically integrable and mixed systems (where integrable and chaotic phase-space regions coexist) within various contexts \cite{Keshavamurthy2011, Tomsovic1994, Podolskiy2003, Backer2008, Backer2011, Relano2008, Steck2001, Frischat1998, Lin1990, Doron1995, Doron2001, Maitra1997}. In such systems, dynamical tunneling was not observed to impact the overall integrability associated with the conserved quantities governing the entire system. In this paper, we use pseudointegrable systems to show that dynamical tunneling can break classical conserved quantities other than energy, giving rise to non-integrable behaviors. 
		
\begin{figure}
			\centering
			\includegraphics[width=1\linewidth]{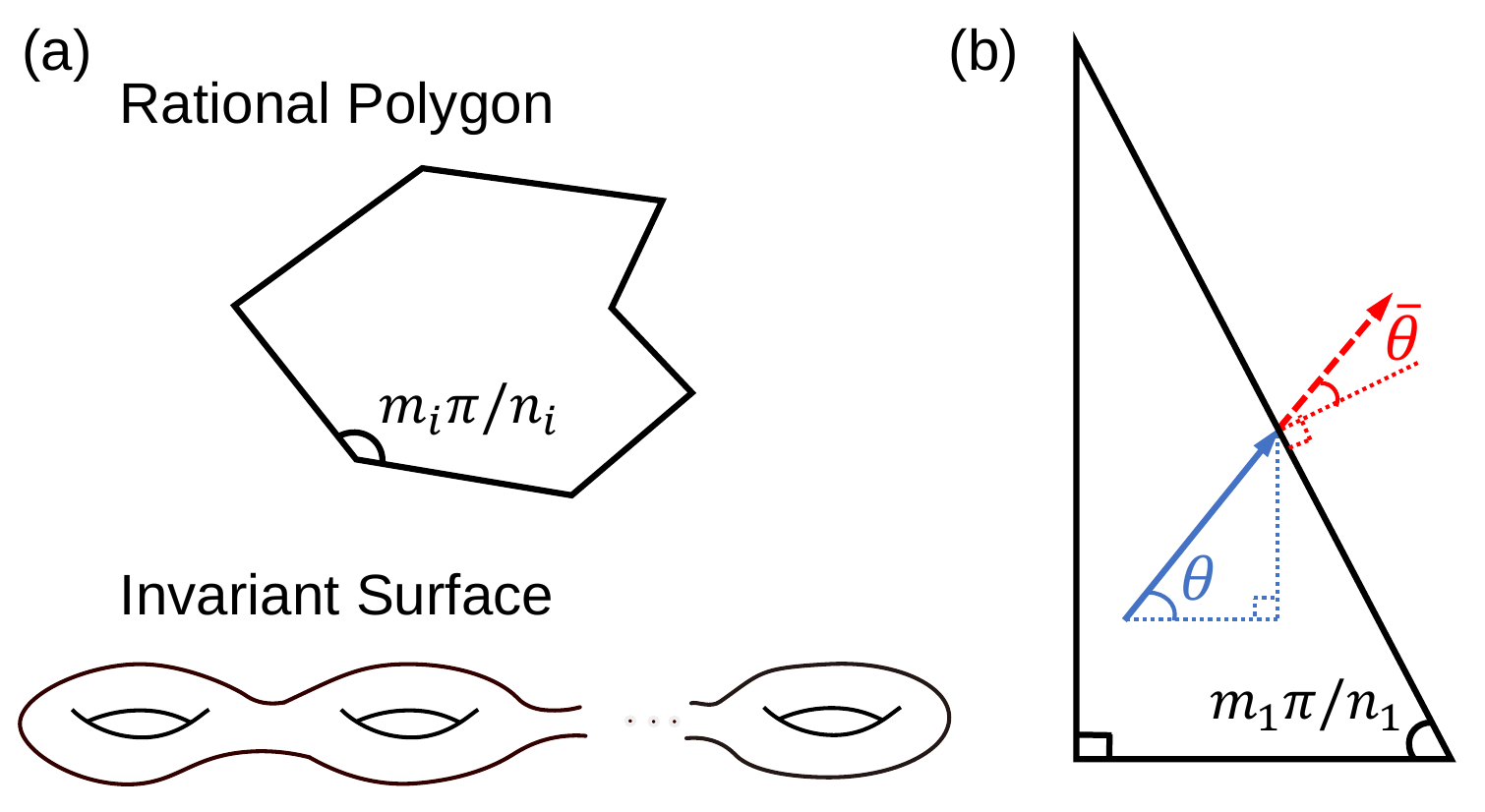}
			\caption{Pseudointegrable systems. (a) Rational polygon and invariant surface $\mathcal{R}$ with constant energy and $T$. Each angle at the vertex $i$  of polygon has a rational degree $m_i\pi/n_i$ with $m_i$ and $n_i$ being coprime integers. At least one vertex has $m_i>1$. (b) Rational right triangle. $m_1$ and $n_1$ are coprime integers. The directions of momentum can be described by two alternative variables: the angle $\theta$ of inclination with respect to the horizontal axis, or the included angle $\bar{\theta}$ with respect to the outer normal vector of sides. }
			\label{fig0}
		\end{figure}

		The pseudointegrable systems were introduced by Richens and Berry in $1981$ \cite{Richens1981}. They are defined as classical Hamiltonian systems with equal degrees of freedom and conserved quantities, where the phase trajectories are restricted to an invariant surface $\mathcal{R}$ featuring a multi-handled sphere with genus larger than $1$, in contrast to the genus-$1$ torus found in integrable systems. Typical examples of such systems are  rational polygon billiards (see Fig.~\ref{fig0}(a)), whose conserved quantity other than energy is given by 
		\begin{equation}\label{eq1}
			T(\cos\theta) \triangleq \cos \left(N\theta\right)
		\end{equation}
		where $\theta$ is the angle between the momentum and the horizontal axis and $N$ is the least common multiple of $n_i$ for all vertices with rational angles $m_i\pi/n_i$ \cite{Gutkin1986}. The conservation arises from the fact that the directions of each classical trajectory follow the orbits of a dihedral group $D_N$.
		
		The dynamical properties of pseudointegrable systems have been studied over decades both classically and quantum mechanically but from very different perspectives. In classical mechanics, people are concerned about how the unremovable singularities on $\mathcal{R}$ split the beams of trajectories \cite{Richens1981, Gutkin1986}. 
		In the quantum counterpart, the exploration has primarily revolved around their energy level statistics \cite{Richens1981,Cheon1989,Shudo1994,Shigehara1994,Wiersig2002}. Following this line, Bogomolny identified a subset of them characterized by the semi-Poisson distribution \cite{Bogomolny1999, Bogomolny2001, Bogomolny2004_1, Bogomolny2001,Bogomolny2004_2, Bogomolny2009, Bogomolny2021}, which had been proposed as a universal intermediate bridging the gap between Poisson statistics and Wigner-Dyson statistics \cite{Shklovskii1993,Kravtsov1994}. Pseudointegrable systems present two advantages for our study: {\it i}) Singularities in phase space were suggested as an important element to provoke dynamical tunneling \cite{Keshavamurthy2011}; {\it ii}) Absence of a classical chaotic sea makes the non-integrable behaviors more attributable to the effects of dynamical tunneling.
		
		Here we focus on the rational right triangle billiards, as depicted in Fig.~\ref{fig0}(b). This subclass has all the features of pseudointegrable systems. We analytically prove the absence of conserved quantity $T$ in their quantum counterparts. The entire proof unfolds in two steps. First, assuming $\hat{T}$ as the quantization of $T$ is conserved, we can expand the eigenstates by $2N$ plane waves with the directions following the orbits of group $D_N$. Second, imposing Dirichlet boundary conditions on the sides, we find that they cannot be satisfied, unless the billiards are completely integrable (see Appendix \ref{Appendix_A} for details). 
		
		To quantify the quantum mechanical breaking of the conserved quantity $T$, we calculate the uncertainty of operator $\hat{T}$ for each eigenstate, defined by
		\begin{equation}\label{eq2}
			\sigma\left(\psi_n\right) \triangleq  \langle \psi_n|\hat{T}^2|\psi_n\rangle - \langle \psi_n|\hat{T}|\psi_n\rangle^2,
		\end{equation}
		where $ \psi_n $ denotes the $n$-th eigenstate. We observe that the decreasing trend of $\sigma\left(\psi_n\right)$ with rising energy cannot be described by elementary functions. Instead, we examine the probability density functions (PDFs) of $\sigma$ within an energy shell, revealing nontrivial distributions that can be accurately fitted by the Brody distribution \cite{Brody1973, Izrailev1990}. The eigenstates with sufficiently small $\sigma$ are collected, and their unfolded level spacing statistics conform to the Poisson distribution. While, the eigenstates with large $\sigma$ showcase a superposition with classical regular orbits with different $T$, regardless of whether they are periodic or not. It is the consequence of dynamical tunneling, which results in avoided crossings among regular states. Inspired by the statistics of $\sigma$, we develop a random matrix model for dynamical tunneling and successfully reproduce the level spacing distributions of pseudointegrable systems.  
		
		\section{conserved quantity} 
		Each classical trajectory has directions being the orbits of group $D_N$, which fold $0\leq \theta \leq 2\pi$ into an interval $[0,\pi/N]$. This corresponds to a quotient map $|\theta\mod (2\pi/N)|$, functioning as a conserved quantity. For simplicity, we use the form \eqref{eq1}. Consequently, the trajectory in phase space can be projected onto $\mathcal{R}$. 
		It is simple to prove $T(\cos\theta)$ is conserved along a classical trajectory, i.e., for a certain reflection at one side, $T(\cos\theta) = \cos \left(N\theta\right) \to \cos \left[N(2j\pi/N \pm \theta)\right] = \cos \left(N\theta\right) = T(\cos\theta)$ for $j = 0,1,\dots, N-1$. Additionally, $T(\cos\theta)$ can be expanded as a polynomial in $\cos\theta$, which can be quantized as the operator $\hat{P}_x/P$, where $P$ is the magnitude of the momentum  and $\hat{P}_x$ is the momentum operator along $x$ direction. 
		\section{Uncertainty} 
		We calculate the quantum right triangle billiards with one interior angle being $\pi/8$, $\pi/5$, $\pi/7$, $2\pi/7$, $\pi/9$, $2\pi/9$ (the angles will be used to represent the corresponding billiards). The genus of $\mathcal{R}$ is given by the formula $1+N\sum_{i}(m_i-1)/(2n_i)$. Consequently, each pair of billiards corresponds to $\mathcal{R}$ with genus-2, 3, 4 . It is noted that the energy level statistics of $\pi/8$ and $\pi/5$ has been found to be semi-Poisson distribution \cite{Bogomolny1999}. For each billiard, $10^5$ eigenstates are calculated using a hybrid of scaling method and decomposition method \cite{Vergini1995, Barnett2001}. We use the Weyl's level $N_{\operatorname{Weyl}}$ to indicate the energy of eigenstates\cite{Baltes1976}, which helps compare different systems with a unified energy scale.
						
\begin{figure}
			\centering
			\includegraphics[width=1\linewidth]{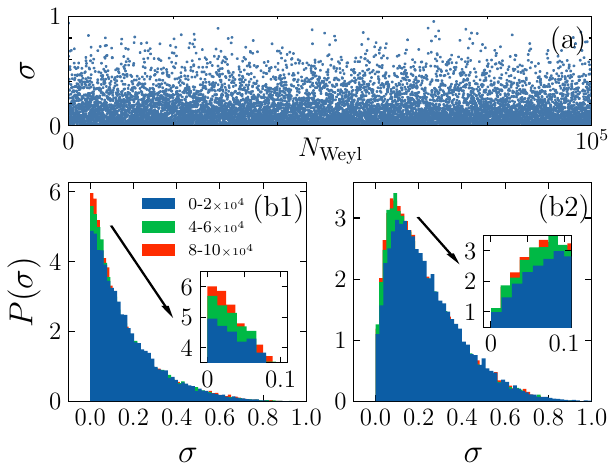}
			\caption{(a) Scatter plots of $\sigma$ in relation to $N_{\operatorname{Weyl}}$ for $\pi/8$ billiard, where $10^4$ points are sampled uniformly. (b1) and (b2) PDFs of $\sigma$. The $10^5$ eigenstates are categorized into three groups based on their energy levels, with the colors representing $N_{\operatorname{Weyl}}$ within the ranges of $0-2\times 10^4$, $4-6\times 10^4$ and $8-10\times 10^4$ in blue, green, and red, respectively. The insets provide a closer look at $\sigma$ smaller than $0.1$. Notably, (b1) and (b2) respectively correspond to $\pi/8$ and $\pi/7$ billiards.}
			\label{fig1}
		\end{figure}

		The direct results of $\sigma$ are depicted in Fig.~\ref{fig1}(a). It demonstrates that most points cluster around the small values of $\sigma$. However, there are still eigenstates with large $\sigma$ values at higher energy levels. Since $\sigma$ does not exhibit a clear decreasing trend with $N_{\operatorname{Weyl}}$,  we analyze $\sigma$ statistically. 
If $\hat{T}$ were conserved, we would expect the PDF of uncertainties, denoted as $P(\sigma)$, to be a delta distribution  at $\sigma = 0$. However, the statistical results show that $P(\sigma)$ follow various nontrivial distributions. This is exemplified by $\pi/8$ and $\pi/7$ billiards in Fig.~\ref{fig1}(b1) and Fig.~\ref{fig1}(b2). Both $P(\sigma)$ have right tails but different peaks. With energy rising, they exhibit a trend to converge to delta distribution. It indicates that in the classical limit, the conserved quantity $T$ will recover. 
		
\begin{figure}
			\centering
			\includegraphics[width=1\linewidth]{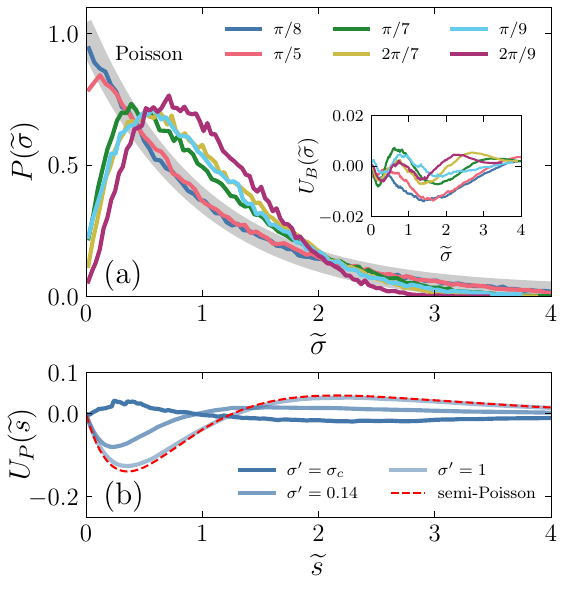}
			\caption{(a) 
				PDFs of $\widetilde{\sigma}$ for $10^5$ eigenstates. The shaded band represents Poisson distribution, i.e., $\exp\left(-\widetilde{\sigma}\right)$. These distributions are fitted by the Brody distribution. The inset displays the differences in CDFs from their optimal fitting distributions, denoted as $U_B(\widetilde{\sigma})$. The optimal fitting parameters $q$ are approximately $1.0$, $1.0$, $1.4$, $1.5$, $1.5$, $1.9$ for $\pi/8$, $\pi/5$, $\pi/7$, $2\pi/7$, $\pi/9$, $2\pi/9$ billiards, respectively. (b) Differences in CDFs of unfolded level spacing with different uncertainty thresholds from the Poisson distribution, denoted as $U_P(\widetilde{s})$. $\sigma'$ is the selected threshold value. For $\pi/8$ billiard, $\sigma_c$ is chosen to be $ 0.0189$ with the number of pseudoregular states amounting to $11481$. Notably, in the absence of uncertainty threshold ($\sigma'=1$), the level spacing statistics for the whole spectrum follows $P_{\operatorname{semi-Poisson}}(\widetilde{s})$, as indicated by the red dashed line. Additionally, the level spacing distribution for the uncertainties less than $0.14$ is shown at the intermediate position between Poisson and semi-Poisson distribution.}
			\label{fig2}
		\end{figure}

		To gain further insights, we consider the unfolded uncertainty, defined as $\widetilde{\sigma}:=\sigma/\operatorname{mean}(\sigma)$, whose domain is extended from $[0,1]$ to $[0,\infty]$. The PDFs of $\widetilde{\sigma}$, denoted as $ P(\widetilde{\sigma})$, are shown in Fig.~\ref{fig2}(a). All $P(\widetilde{\sigma})$ presented here can be well fitted by Brody distributions $P_{\operatorname{Brody}}(\widetilde{\sigma}) = a(q+1)\widetilde{\sigma}^q\exp\left(-a\widetilde{\sigma}^{q+1}\right)$, $a = \Gamma\left[(q+2)/(q+1)\right]^{q+1}$ with a single parameter $q\in[0,1]$. In particular, for $\pi/8$ and $\pi/5$ billiards, whose level statistics are semi-Poissonian, $P(\widetilde{\sigma})$ closely resemble the $q=0$ Poisson distribution $\exp\left(-\widetilde{\sigma}\right)$. The differences in cumulative distribution functions (CDFs) from the best-fitted model, denoted by $U_B(\widetilde{\sigma}):= \int_0^{\widetilde{\sigma}} d\widetilde{\sigma}_1 \left[P(\widetilde{\sigma}_1) - P_{\operatorname{Brody}}^{*}(\widetilde{\sigma}_1)\right]$ where the superscript $*$ represents ``optimal fitting", are shown in the inset, which give the errors less than $2\%$. This result shows that $P(\widetilde{\sigma})$ for genus-3, 4 billiards have a polynomial repulsion at small $\widetilde{\sigma}$ and a super-exponential tail at large $\widetilde{\sigma}$. Roughly,	$P(\widetilde{\sigma})$ exhibits an elevated $q$ with genus increasing, as shown in Fig.~\ref{fig2}(a). This may be understood from Diophantine approximations. Consider a right triangle billiard, as depicted in Fig.~\ref{fig0}(b), where the ratio $m_1/n_1$ is a good Diophantine approximation of some irrational number. Then, $N$ is typically large, resulting in a high genus. Since irrational triangle billiards lack the conserved quantity $T$, a $\mathcal{R}$ with higher genus naturally implies a more chaotic system. 
	
		To confirm our expectation that the level statistics of eigenstates with small enough $\sigma$ follows a Poisson distribution, we scan the values of $\sigma$ to determine a small enough threshold $\sigma_c$, such that the unfolded level spacing distribution of eigenstates with $\sigma < \sigma_c$ aligns with the Poisson distribution with minimal fitting error. This can be formulated as $ P\left(\widetilde{s} \;| \left\{\psi_n,\sigma(\psi_n)<\sigma_c\right\}\right) \simeq P_{\operatorname{Poisson}}\left(\widetilde{s}\right)$ where $P_{\operatorname{Poisson}}(\widetilde{s}) = \exp\left(-\widetilde{s}\right)$ and $\widetilde{s}$ is the unfolded level spacing. We call the eigenstates with $\sigma < \sigma_c$ as ``pseudoregular" states. The results for $\pi/8$ billiard are depicted in Fig.~\ref{fig2}(b), where the differences in CDFs of unfolded level spacing with different thresholds $\sigma'$ from the Poisson distribution are calculated, defined as $U_P(\widetilde{s}):= \int_0^{\widetilde{s}} d\widetilde{s}_1 \left[P\left(\widetilde{s}_1 \;| \left\{\psi_n,\sigma(\psi_n)<\sigma'\right\}\right) - P_{\operatorname{Poisson}}(\widetilde{s}_1)\right]$. It shows that the level statistics	of pseudoregular states approximates to $P_{\operatorname{Poisson}}(\widetilde{s})$, implying an absence of level repulsions in the spectrum if $\hat{T}$ is conserved. It is established that when all the eigenstates are involved ($\sigma' = 1$), the level spacing statistics for the whole spectrum follows a semi-Poisson distribution, i.e., $P_{\operatorname{semi-Poisson}}(\widetilde{s}) = 4\widetilde{s}\exp\left(-2\widetilde{s}\right)$. By increasing the threshold to $\sigma' = 0.14 > \sigma_c$, the level statistics occupies an intermediate position between the Poisson and semi-Poisson distribution. This demonstrates that by controlling the uncertainty threshold $\sigma'$, the level statistics can be changed between these two distributions. We will see that the eigenstates with $\sigma>\sigma_c$ are characterized by the superposition of classical trajectories with distinct values of $T$, which is caused by dynamical tunneling.
		\begin{figure}
			\centering
			\includegraphics[width=1\linewidth]{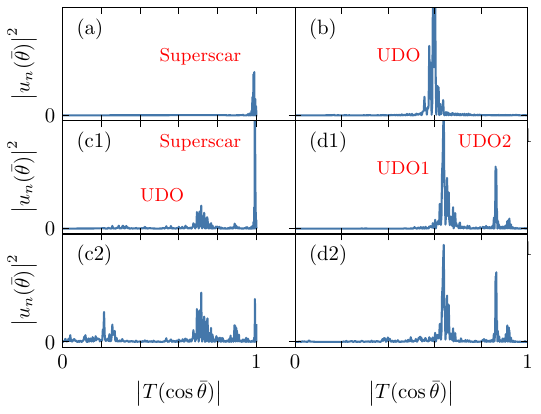}
			\caption{Profiles of different types of eigenstates for $\pi/8$ billiard, illustrated by $u_n(\bar{\theta})$ with respect to $\left|T(\cos \bar{\theta})\right|$. (a) Superscar state with $N_{\operatorname{Weyl}}\approx10017.3$ and $\sigma\approx 0.006$. (b) Spacial uniformly distributed orbit with $N_{\operatorname{Weyl}}\approx40090.0$ and $\sigma\approx 0.002$. (c1) Superposition of a single superscar and a single UDO with $N_{\operatorname{Weyl}}\approx60144.6$ and $\sigma\approx 0.563$. (c2) is the eigenstate next to (c1) with $N_{\operatorname{Weyl}}\approx60144.1$ and $\sigma\approx 0.489$. (d1) Superposition of two UDOs with $N_{\operatorname{Weyl}}\approx61476.4$ and $\sigma\approx 0.461$. (d2) is the eigenstate next to (d1) with $N_{\operatorname{Weyl}}\approx61476.6$ and $\sigma\approx 0.509$.}
			\label{fig3}
		\end{figure}		
	
		\section{Dynamical Tunneling} The direct approach to investigate dynamical tunneling is mapping the eigenstates into phase space \cite{Wigner1932, Husimi1940, Wang2021}. 
Fortunately, in pseudointegrable systems, each classical orbit has a specific value of $T$. As a result, each eigenstate that may be
a superposition of different classical orbits 
can be converted into a distribution of different $T$ values . 
This distribution is expressed by the boundary function $u_n(\bar{\theta})$ of $\psi_n$, where $\bar{\theta}$ is the included angle between the outer normal vector of sides and the momentum as shown in Fig.~\ref{fig0}(b). 
This method follows Ref.\cite{Backer1998} and its details are given in Appendix \ref{Appendix_B}. With $\bar{\theta}$, the conserved quantity $T$ can be rewritten as $T =|T(\cos \bar{\theta})| = |\cos\left(N\bar{\theta}\right)|$, where the absolute value arises at the hypotenuse $\bar{\theta} = \pi/2+m\pi/n-\theta$. In this way, $\bar{\theta}$ becomes the connection between $u_n\left(\bar{\theta}\right)$ and $T$. 
		
		The results of $\pi/8$ billiard are shown in Fig.~\ref{fig3}. For this billiard, there are only two types of classical orbits: periodic orbits (POs) and uniformly distributed orbits (UDOs) that cover the entire billiard table \cite{Veech1989}. It has been established that, in rational right triangle billiards, nearly every periodic orbit contains segments that are perpendicular to a specific side. Consequently, almost all POs has $T= 1$, while UDOs have various values of $T$. Quantum mechanically, there exists ``superscar" states, named by Bogomolny \cite{Bogomolny2004_1, Bogomolny2006, Aberg2008}, exhibiting a superposition of spatially parallel POs. $T$ cannot be used to distinguish different superscar states as they all equal to $1$.
		
		In both Fig.~\ref{fig3}.(a) and Fig.~\ref{fig3}.(b), $\left|u_n(\bar{\theta})\right|^2$ exhibit a single peak, illustrating the features of superscar and UDO, respectively. Both of them are pseudoregular states, characterized by uncertainties $\sigma$ smaller than $\sigma_c$. However, in Fig.~\ref{fig3}.(c1) and Fig.~\ref{fig3}.(d1), orbits with different values of $T$ are superposed (one superscar plus one UDO in (c1), two UDOs in (d1)), causing their relatively large $\sigma$. Notably, eigenstates with small wavenumbers may distribute with a considerable width on the values of $T$, leading to an elevated $\sigma$. But the presence of separated peaks here eliminates this concern. The eigestates neighboring to Fig.~\ref{fig3}.(c1) and Fig.~\ref{fig3}.(d1) are shown in Fig.~\ref{fig3}.(c2) and Fig.~\ref{fig3}.(d2). And each pair of them shares common peak positions, signifying common tunneled components. Additionally, their individual level splittings are much smaller than the mean level spacing $1$. This is another indication of dynamical tunneling. In Fig.~\ref{fig3}.(c2), there emerges sub-peaks at small $T$ comparing to Fig.~\ref{fig3}.(c1), which may illustrate more resonant orbits. Therefore, the results indicate that dynamical tunneling destroys the conservation of $\hat{T}$ by superposing trajectories with different directions.
		
		\section{Avoided Crossing} 
		The dynamical tunneling can be modeled by a random matrix Hamiltonian
		\begin{equation}\label{eq4}
			h = 
			\begin{pmatrix}
				\epsilon & \gamma\\
				\gamma & -\epsilon
			\end{pmatrix}\,,
		\end{equation}
		where the diagonals $\pm \epsilon$ are 
		energy levels of different classical orbits, and 
		the off-diagonal $\gamma$ represent tunneling rate between classical orbits with different values of $T$. 
		$\gamma$ is real due to the time-reversal symmetry.
As different classical orbits are not correlated, it is reasonable
to assume that $\epsilon$ has a Poisson distribution, i.e.,   
$P(\epsilon) = \exp\left(-2|\epsilon|\right)$. 

\begin{figure}
			\centering
			\includegraphics[width=1\linewidth]{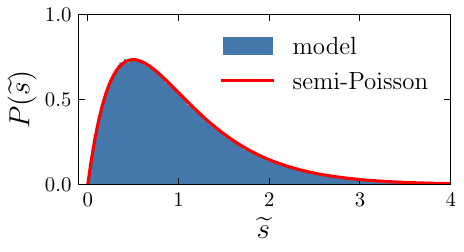}
			\caption{Unfolded level spacing distribution of model \eqref{eq4} using the tunneling rate \eqref{eq5}. The level spacings are collected from $10^6$ ensembles. The red line represents the semi-Poisson distribution.}
			\label{fig4}
		\end{figure}

The numerical results in Fig.~\ref{fig1}(b1) 
and Fig.~\ref{fig2}(a) suggest that the distribution $g(\gamma)$ of 
$\gamma$  have the same shape as $P(\sigma)$, that is, 
\begin{equation}\label{eq5}
			g(\gamma)=g(-\gamma)\approx  \frac{1}{2}\exp\left(-{|\gamma|}\right).
		\end{equation}
Once the two distributions $P(\epsilon)$ and $g(\gamma)$ are 
known, we can compute the level spacing distribution of \eqref{eq4}, which is compared with $P_{\operatorname{semi-Poisson}}(\widetilde{s})$ in Fig.~\ref{fig4}, demonstrating a perfect fit. Therefore, combing the results in Fig.~\ref{fig3}(b) and Fig.~\ref{fig4}, we can conclude that dynamical tunneling is responsible for distorting the level spacing distribution from a Poisson distribution to the semi-Poisson distribution. We expect this mechanism applies to all pseudointegrable systems with diverse level spacing distributions.	
 
		\section{Discussion} 			
		The dynamical tunneling observed in pseudointegrable systems shows different effects compared to those in integrable and mixed systems. We break down them into three aspects. 
{\it i}) Conserved quantity.  In mixed systems, one can define a ``local" conserved quantity for the trajectories within symmetry-related integrable islands. Dynamical tunneling among these integrable islands or from them into the chaotic sea can destroy this ``local" conserved quantity \cite{Keshavamurthy2011, Tomsovic1994, Podolskiy2003, Backer2008, Backer2011, Relano2008, Steck2001, Frischat1998, Lin1990, Doron1995, Doron2001}. However, here, dynamical tunneling has the potential to break the conserved quantity $T$ governing the entire systems by superposing classical orbits with different values of $T$ (as shown in Fig.~\ref{fig1} and Fig.~\ref{fig3}).	
{\it ii}) Level statistics. For mixed systems, dynamical tunneling can only introduce 
		sub-leading corrections to the Berry-Robnik level statistics, since it weakly couples regular and chaotic states, thus increasing small distances between the corresponding levels \cite{Berry1984, Tomsovic1994, Backer2008, Backer2011, Podolskiy2003, Relano2008}. The level repulsion in pseudointegrable systems is completely induced by dynamical tunneling (as shown in Fig.~\ref{fig2}.(b)). {\it iii}) Tunneling rate. In principle, the tunneling rate between two resonant ``double-well states" decreases exponentially with $\hbar^{-1}$, which can be expressed as $\ln\gamma \propto -\xi/\hbar$ where $\xi$ is a parameter related to systems and resonant orbits (for a double-well system , $\xi$ is the imaginary part of action over the energy barrier in the middle) \cite{Tomsovic1994, Keshavamurthy2011}. In the presence of more resonant states, the tunneling rates between the original pair can be enhanced or suppressed by several orders of magnitude at certain values of $\hbar^{-1}$ \cite{Steck2001, Frischat1998, Lin1990, Doron1995, Doron2001}. We  formulate this qualitatively as $\xi = \xi(\hbar^{-1}) = \bar{\xi} + \sum_{i}\lambda_i\delta(\hbar^{-1} - \hbar_i^{-1})$ where $\bar{\xi}$ is the average value over a shell of $\hbar^{-1}$, $\hbar_i^{-1}$ is the enhanced (or suppressed) point and $\lambda_i$ is the corresponding strength. In the limit of strong tunneling, the distribution of $\hbar_i^{-1}$ becomes dense on the shell, allowing us to treat $\xi$ as a random variable centered around $\bar{\xi}$. This is consistent	with our semi-empirical distribution of tunneling rate \eqref{eq5}. However, a rigorous quantitative derivation of tunneling rate remains an open problem.
		
		\section*{ACKNOWLEDGMENTS}
		LQK and BW are supported by National Natural Science Foundation of China 
		(Grant No. 11921005 \& 92365202), and Shanghai Municipal Science and Technology Major Project (Grant No.2019SHZDZX01). ZG is supported by The University of Tokyo Excellent Young Researcher Program.

			\appendix
			\renewcommand\theequation{\thesection\arabic{equation}} 
			
			\section{A proof of conserved quantity breaking in rational right triangle billiards} \label{Appendix_A}
			
\begin{figure}
				\centering
				\includegraphics[width=1\linewidth]{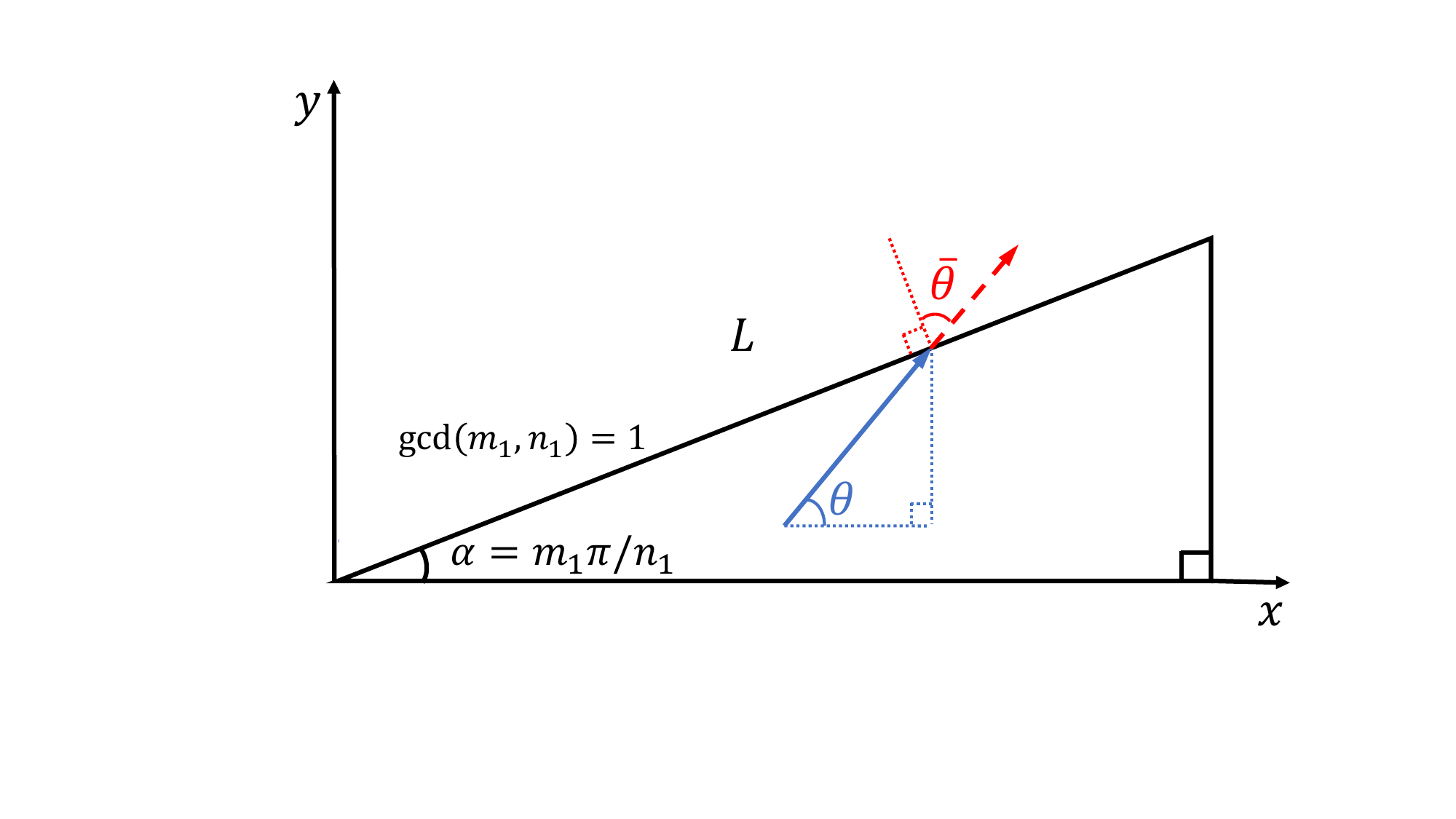}
				\caption{Rational right triangle billiard. This billiard includes at least one vertex with angle $\alpha$ being $m_1\pi/n_1$ where $n_1$ is even,  $m_1$ ($m_1\neq 1$) and $n_1$ are coprime integers. $L$ is the length of hypotenuse. The directions of billiard can be described by either of two alternative variables: the angle $\theta$ of inclination with respect to the $x$-axis, or the included angle $\bar{\theta}$ with respect to the outer normal vector of sides.}
				\label{SM_fig0}
			\end{figure}

			Consider a rational right triangle billiard described in Fig.~\ref{SM_fig0}. The direction $\theta$ can undergo three kinds of reflections by sides: $\theta \to -\theta$ (reflection at $y=0$), $\theta \to \pi-\theta$ (reflection at $x=L\cos\alpha$) and $\theta \to 2\alpha - \theta$ (reflection at $y = x\tan\alpha$). These can generate a dihedral group $D_{N}$ ($N=n_1$) with dimension $2N$, yielding a conserved quantity $T = \cos N\theta$ independent with the energy. $T$ can be quantized after expanding to the polynomial of $\cos \theta = {P}_x/P$ where $P_x$ is the momentum along $x$-direction and $P$ is the norm of total momentum.
			
			The conserved quantity $T$ originates from the existence of quotient projection mapping each classical trajectory on the surface $S^1/D_N$, where $S^1$ is the unit circle. It means that for a classical trajectory, the possible directions, denoted as $\pm\theta_{j}$, can be related to a reference $\theta_0 \in [0,\pi/N]$ so that  
			\begin{equation}\label{eqSM1}
				\theta_j = \theta_0 + \frac{2j\pi}{N}, \quad  j = 0,1,\dots,N-1.
			\end{equation}
			If the quantized operator $\hat{T}$ remains conserved in quantum mechanics, the eigenstate should also follow the orbits of group $D_N$, just as each classical trajectory does.
			In this manner, the eigenstate, denoted as $|\psi\rangle$, can be expanded by plane waves with direction $\pm\theta_{j}$,
			\begin{equation}\label{eqSM2}
				|\psi\rangle=\sum_{j=0}^{N-1}\left(c_j^{+}\left|k \cos \theta_j, k \sin \theta_j\right\rangle+c_j^{-}\left|k \cos \theta_j,-k \sin \theta_j\right\rangle\right)
			\end{equation} 
			where $c_j^{ \pm}$ are the superposing parameters. Then, the Dirichlet boundary conditions at $y = 0$, $y = x\tan\alpha$ and $x=L\cos\alpha$ can be evaluated sequentially:
			\begin{align}
				\sum_{j=0}^{N-1}\left(c_j^{+}+c_j^{-}\right) e^{ik x \cos \theta_j}=0, \label{eqSM3}\\
				\sum_{j=0}^{N-1}\left[c_j^{+} e^{ik r \cos \left(\theta_j-\alpha\right)}+c_j^{-} e^{ik r \cos \left(\theta_j+\alpha\right)}\right]=0, \label{eqSM4}\\
				\sum_{j=0}^{N-1}\left(c_j^{+} e^{ik y \sin \theta_j}+c_j^{-} e^{-ik y \sin \theta_j}\right) e^{ik L \cos \alpha \cos \theta_j}=0, \label{eqSM5}
			\end{align}
			where $r=\sqrt{x^2+y^2} \in$ $[0, L]$. Suppose the subscript of $c_j^{\pm}$ has a cyclic relation $c_j^{ \pm}:=c_{j \bmod N}^{ \pm}$ for $j\in\mathbb{Z}$, Eq.~\eqref{eqSM4} can be rewritten as
			\begin{equation}\label{eqSM6}
				\sum_{j=0}^{N-1}\left(c_{j+m}^{+}+c_j^{-}\right) e^{ik r \cos \left(\theta_j+\alpha\right)}=0.
			\end{equation}
			
			We can prove that Eq.~\eqref{eqSM3} and Eq.~\eqref{eqSM6} are equivalent to 
			\begin{equation}\label{eqSM7}
				c_j^{+}+c_j^{-}=c_{j+m}^{+}+c_j^{-}=0, \quad j=0,1, \ldots, N-1.
			\end{equation}
			We take the $i$-th $(i=1,2, \ldots, N-1)$ derivative of the both sides of Eq.~\eqref{eqSM3} and make $x$ equal to zero. In this way, we obtain $N$ homogeneous linear equations by defining $a_j:=c_j^{+}+c_j^{-}$,
			\begin{equation}\label{eqSM8}
				\left[\begin{array}{ccccc}
					1 & 1  & \ldots & 1 \\
					\cos\theta_0 & \cos\theta_1  & \ldots & \cos\theta_{N-1} \\
					\cos^2\theta_0 & \cos^2\theta_1   & \ldots & \cos^{2}\theta_{N-1} \\
					\vdots  & \vdots & \ddots & \vdots \\
					\cos^{N-1}\theta_{0} & \cos^{N-1}\theta_{1}  & \ldots & \cos^{N-1}\theta_{N-1}
				\end{array}\right] \left[\begin{array}{c}
					a_0 \\
					a_1 \\
					\vdots \\
					a_{N-1}
				\end{array}\right]=0.
			\end{equation}    
			Its determinant is known as Vandermonde determinant, yielding
			\begin{equation}\label{eqSM9}
				\operatorname{det}\left[\cos ^{i-1} \theta_j\right]_{N \times N}=\prod_{0 \leq i<j \leq N-1}\left(\cos \theta_j-\cos \theta_i\right).
			\end{equation}   
			If let $\cos \theta_j = \cos \theta_i$, we obtain $\theta_0=({N-i-j} )\pi/{N} $ which cannot be confined in the domain $\left(0, {\pi}/{N}\right)$. As a result, this determinant is nonzero for $\theta_{0}\in\left(0, {\pi}/{N}\right)$. In this way, the solution to Eq.~\eqref{eqSM8} is a null vector, i.e.,
			\begin{equation}\label{eqSM10}
				a_j=c_j^{+}+c_j^{-}=0.
			\end{equation}
			When $\theta_{0} = 0, \pi/N$, the billiard has $N$ directions corresponding to degenerate cases with $\theta_{j} = -\theta_{N-j}$. We redefine the superposing parameter as $c_j$, so that 
			$|\psi\rangle=\sum_{j=0}^{N-1}c_j\left|k \cos \theta_j, k \sin \theta_j\right\rangle$. 
			For $\theta_{0} = 0$, the degenerate version of boundary conditions Eq.~\eqref{eqSM3} and Eq.~\eqref{eqSM4} become
			\begin{align}
				c_0 e^{ikx} + c_{N/2}e^{-ikx} + \sum_{j=1}^{N/2-1}\left(c_j+c_{N-j}\right) e^{ikx \cos \theta_j}=0, \label{eqSM11}\\
				\sum_{j=0}^{(m-1)/2}(c_j + c_{m-j}) e^{ik r \cos \left(\theta_j-\alpha\right)}+\nonumber\\
				\sum_{j=m+1}^{(m+N-1)/2}(c_j + c_{m+N-j}) e^{ik r \cos \left(\theta_j-\alpha\right)}=0. \label{eqSM12}
			\end{align}
			These two relations are equivalent to $	c_j+c_{N-j}=c_j+c_{m-j}=0 $, yielding $c_{m-j} = c_{-j}$. Consequently, all $c_j$ must equal to $c_0=0$.
			For $\theta_{0} = \pi/N$, we have 
			\begin{align}
				\sum_{j=0}^{N / 2-1}\left(c_j+c_{N-1-j}\right) e^{ik x \cos \theta_j}=0, \label{eq13}\\
				c_{(m-1) / 2} e^{ik r}+c_{(m+N-1) / 2} e^{-ik r}+ \nonumber\\
				\sum_{j=0}^{(m-3) / 2}\left(c_j+c_{m-1-j}\right) e^{ik r \cos \left(\theta_j-\alpha\right)}+ \nonumber\\
				\sum_{j=m}^{(m+N-3) / 2}\left(c_j+c_{m+N-1-j}\right) e^{ik r \cos \left(\theta_j-\alpha\right)}=0. \label{eqSM14}
			\end{align}
			They are equivalent to $c_j+c_{N-j-1} = c_j + c_{m-j-1} = 0$, yielding $c_{m-j-1} = c_{1-j}$. Consequently, all $c_j$ must equal to $c_{(m-1)/2}=0$. So far, we have proved an absence of common eigenstates for the degenerate cases of $\theta_{0} = 0, \pi/N$.
			
			Using the same tricks on Eq.~\eqref{eqSM6}, we can obtain $c_{j+m}^++c_j^-=0$.
			
			Applying the relation Eq.~\eqref{eqSM7} on the eigenstates $|\psi\rangle$, we get
			\begin{equation}\label{eqSM15}
				\psi(x,y) = \sum_{j=0}^{N / 2-1} \sin \left(k x \cos \theta_j\right) \sin \left(k y \sin \theta_j\right).
			\end{equation}
			One can verify $\psi(x,y)$ is invariant under the action of dihedral group $D_N$, $\theta_j \to 2\beta \pi/N\pm\theta_j$ for $\beta\in \mathbb{Z}$. $\psi(x,y)$ is a irreducible representation of group $D_N$.
			
			We apply the boundary condition Eq.~\eqref{eqSM5} to Eq.~\eqref{eqSM15}, yielding 
			\begin{equation}\label{eqSM16}
				\sum_{j=0}^{N / 2-1} \sin \left(k L \cos \alpha \cos \theta_j\right) \sin \left(k y \sin \theta_j\right)=0.
			\end{equation}
			We can prove that Eq.~\eqref{eqSM16} is equivalent to 
			\begin{equation}\label{eqSM17}
				\sin \left(k L \cos \alpha \cos \theta_j\right)=0 \quad \text{or} \quad k L \cos \alpha \cos \theta_j=M_j \pi
			\end{equation}
			where $M_j \in \mathbb{Z}, \forall j=0,1, \ldots, N / 2-1$. This proof is similar to the proof of Eq.~\eqref{eqSM7}.
			We take the $2 i-1$-th $(i=1,2, \ldots, N / 2)$ deviatives of both side of Eq.~\eqref{eqSM11} and make $y$ equal to zero. We obtain $N-1$ homogeneous linear equations by defining $b_j:=\sin \left(k L \cos \alpha \cos \theta_j\right) \sin \theta_j$, 
			\begin{equation}\label{eqSM18}
				\left[\begin{array}{ccccc}
					1 & 1  & \ldots & 1 \\
					\sin ^2 \theta_0 & \sin ^2 \theta_1  & \ldots & \sin ^2 \theta_{N-1} \\
					\sin ^4 \theta_0 & \sin ^4 \theta_1   & \ldots & \sin ^4 \theta_{N-1} \\
					\vdots  & \vdots & \ddots & \vdots \\
					\sin ^{N-2} \theta_{0} & \sin ^{N-2} \theta_{1}  & \ldots & \sin ^{N-2} \theta_{N-1}
				\end{array}\right] \left[\begin{array}{c}
					b_0 \\
					b_1 \\
					\vdots \\
					b_{N-1}
				\end{array}\right]=0.
			\end{equation}  
			Its determinant is nonzero, so the solution is $b_j = 0$, yielding $\sin \left(K L \cos \alpha \cos \theta_j\right) = 0$. 
			
			The relation Eq.~\eqref{eqSM17} can be used as the quantization condition. We divide it by itself with a different subscript, and obtain
			\begin{equation}\label{eqSM19}
				\frac{\cos \theta_j}{\cos \theta_{j^{\prime}}}=\frac{M_j}{M_{j^{\prime}}} \in \mathbb{Q}, \quad  j, j^{\prime}=0,1, \ldots, N-1
			\end{equation}
			where $M_j=-M_{N-j-1}$ for $ j=N / 2, N / 2+1, \ldots, N-1$. By using the identity
			\begin{equation}\label{eqSM20}
				\cos \theta_j+\cos \theta_{j^{\prime}}=2 \cos \frac{\theta_j+\theta_{j^{\prime}}}{2} \cos \frac{\theta_j-\theta_{j^{\prime}}}{2}
			\end{equation}
			and setting $j^{\prime}$ to be $j+2$, we find that
			\begin{equation}\label{eqSM21}
				2 \cos \frac{2 \pi}{N}=\frac{\cos \theta_j}{\cos \theta_{j+1}}+\frac{\cos \theta_{j+2}}{\cos \theta_{j+1}} \in \mathbb{Q},
			\end{equation}
			illustrating the l.h.s. is rational. It is known that if the cosine of a rational angle is a rational number, the only possible cosine values are $\pm 1, \pm 1 / 2$ and $0$ (A proof was provided in \cite{Jahnel2010}). They correspond to the right triangle billiards with $\alpha = \pi/6$ and $\pi/4$, which are the only two complete integrable cases. For others, Eq.~\eqref{eqSM21} are not satisfied. Therefore, we conclude that in rational right triangle billiards, $\hat{T}$ is not conserved unless the billiards are integrable.
			
			\section{Poincar\'e-Husimi representation} \label{Appendix_B}
			Poincar\'e-Husimi representation is used to compare the eigenstates with classical trajectories \cite{Backer1998, Lozej2022}. In this representation, the Poincar\'e section is typically chosen at the boundary of billiards, yielding the boundary function defined as the normal derivative of eigenfunction, i.e.,
			\begin{equation}\label{eqSM22}
				u'_n(s)\triangleq \hat{\mathbf{n}}(s) \cdot \nabla\psi_n(x(s), y(s)),
			\end{equation}
			where $(x(s), y(s))$ is a point on the boundary parameterized by the arc-length $s$ and $\hat{\mathbf{n}}(s)$ denotes the outer normal unit vector at $(x(s), y(s))$. Conversely, the wave function can be obtained by boundary integral
			\begin{equation}\label{eqSM23}
				\psi_n(x,y) = -\oint_{\partial B} ds \;u'_n(s) G[x,y;x(s),y(s)],
			\end{equation}
			where $\partial B$ is the billiard boundary curve, $(x,y)$ is the point inside the billiard table and $G[x,y;x(s),y(s)]$ is the 2-dimensional free particle Green's function.
			In phase space, the conjugate variable to $s$ is $\sin\bar{\theta}$, where $\bar{\theta}$ represents the included angle between the outer normal vector of sides and the velocity of billiard flow, as shown in Fig.~\ref{SM_fig0}. The conjugate function $u_n(\bar{\theta})$ is expressed as
			
			\begin{equation}\label{eqSM24}
				u_n(\bar{\theta})= \oint_{\partial B} ds \;\exp \left(i k s\sin\bar{\theta}\right) u'_n(s),
			\end{equation} 
			where $k$ is the wavenumber. 
			It should be emphasized that the relations between $\theta$ and $\bar{\theta}$ have three kind depending on the sides: $\bar{\theta} = \theta + \pi/2$ (for $y=0$), $\bar{\theta} = \theta$ (for $x=L\cos\alpha$), $\bar{\theta} = \pi/2+\alpha-\theta$ (for $y=x\tan\alpha$). Hence, $|T(\cos \bar{\theta})|$ remains the classical conserved quantity. Without ambiguity, we also use the symbol $T$ to denote $|T(\cos \bar{\theta})|$. The values of $T$ can be used to distinguish different trajectories. For example, in $\pi/8$ billiard, there are only two types of classical orbits: periodic orbits (POs) and uniformly distributed orbits (UDOs) that cover the entire billiard table \cite{Veech1989}. It has been established that, almost each PO possesses segments perpendicular to side of right triangle. Consequently, almost all POs has $T= 1$, while UDOs have various values of $T$. Quantum mechanically, there exists ``superscar" states, exhibiting a superposition of spatially parallel POs \cite{Bogomolny2004_1, Bogomolny2006, Aberg2008}. $T$ cannot be used to distinguish different superscar states as they all equal to $1$. But, the complete details of trajectories can be obtained by projecting the eigenstate using Husimi functions on the phase space expanded by $s$ and $\sin\bar{\theta}$.
			
			The Husimi function $h(q,p)$ with $q = s$, $p=\sin\bar{\theta}$ is defined as
			\begin{equation}\label{eqSM25}
				h(q, p) \triangleq \frac{1}{A}\left|\oint_{\partial B} c_{(q, p)}^{k}(l) u'_n(l) d l\right|^2,
			\end{equation}
			where $A$ is the normalization parameter. $c_{(q, p)}^{k}(l)$ is the standard coherent state, i.e.,
			\begin{equation}\label{eqSM26}
				\begin{aligned}
					c_{(q, p)}^k(l) = & \sum_{j \in \mathbb{Z}} \exp [i k p(l-q+j \mathcal{L})] \\
					& \times \exp \left[-\frac{k}{2}(l-q+j \mathcal{L})^2\right]
				\end{aligned}
			\end{equation}
			where $\mathcal{L}$ is the length of boundary and the sum of $j$ ensures the coherent states being periodic	with a period of $\mathcal{L}$.
			
			\begin{figure}
				\centering
				\includegraphics[width=1\linewidth]{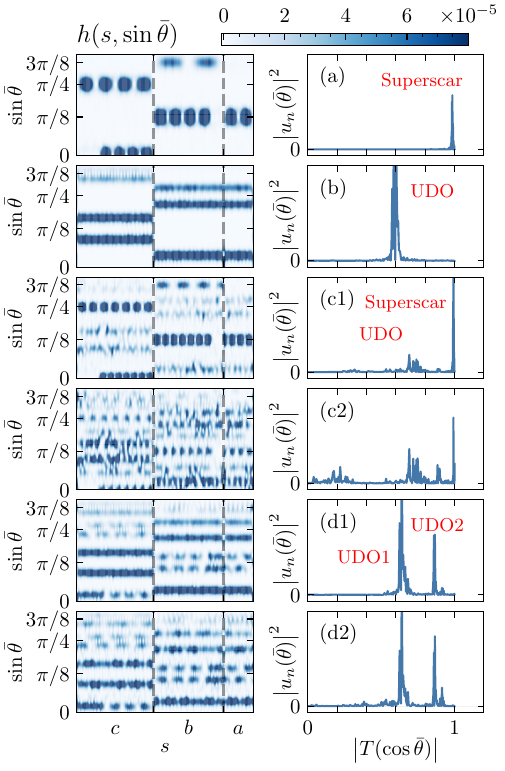}
				\caption{Trajectories of typical eigenstates for $\pi/8$ billiard. The left column are the distributions of Husimi functions $h(s,\sin\bar{\theta})$ on the phase space spanned by $s$ and $\sin\bar{\theta}$. $c$, $b$ and $a$ respectively represent the hypotenuse, the adjacent side and the opposite side of $\pi/8$ corner. The right column are the conjugate boundary function $u_n(\bar{\theta})$ with respect to the conserved quantity $\left|T(\cos \bar{\theta})\right|$. Each row corresponds to the same eigenstate.	(a) A superscar state with $N_{\operatorname{Weyl}}\approx10017.3$ and $\sigma\approx 0.006$, typically exhibiting a superposition of parallel periodic orbits. (b) A spacial uniformly distributed orbit with $N_{\operatorname{Weyl}}\approx40090.0$ and $\sigma\approx 0.002$. (c1) A superposition of a single superscar and a single UDO with $N_{\operatorname{Weyl}}\approx60144.6$ and $\sigma\approx 0.563$. (c2) is the eigenstate next to (c1) with $N_{\operatorname{Weyl}}\approx60144.1$ and $\sigma\approx 0.489$. (d1) A superposition of two UDOs with $N_{\operatorname{Weyl}}\approx61476.4$ and $\sigma\approx 0.461$. (d2) is the eigenstate next to (d1) with $N_{\operatorname{Weyl}}\approx61476.6$ and $\sigma\approx 0.509$.}
				\label{SM_fig1}
			\end{figure}
			
			The results for $\pi/8$ billiard are shown in Fig.~\ref{SM_fig1}. And the Poincar\'e-Husimi representations and the plot of the conjugate boundary function $u_n(\bar{\theta})$ with respect to the conserved quantity $T$ are compared to illustrate the trajectories of typical eigenstates with different $\sigma$ (the uncertainty of $\hat{T}$). The eigenstate in Fig.~\ref{SM_fig1}.(a) is a superscar state, superposed by two parallel POs with segments perpendicular to the hypotenuse. The eigenstate in Fig.~\ref{SM_fig1}.(b) illustrates a UDO. Both of these are pseudoregular states, each exhibiting a single peak at the corresponding value of $T$ and thus possessing the conserved quantity $\hat{T}$. 
			The eigenstate in Fig.~\ref{SM_fig1}.(c1) is a superposition of one superscar (including four parallel POs) and one UDO from the observation on $h(s, \sin\bar{\theta})$. Consequently, it has two separated peaks on the distribution of $T$, causing an elevated $\sigma$. The eigenstate in Fig.~\ref{SM_fig1}.(c2), which is next to the one in Fig.~\ref{SM_fig1}.(c1), includes the trajectories in Fig.~\ref{SM_fig1}.(c1) plus an additional UDO (this UDO also exists in Fig.~\ref{SM_fig1}.(c1) but with a minimal peak). The eigenstates in Fig.~\ref{SM_fig1}.(d1) and Fig.~\ref{SM_fig1}.(d2) are another pair with the energy splitting much smaller than the mean level spacing and exhibit a dynamical tunneling between two UDOs regular state.

		\end{CJK*}

\end{document}